\documentclass[aps,twocolumn,groupedaddress,superscriptaddress,showpacs,prl,floatfix]{revtex4}

\usepackage{graphicx}
\usepackage{dcolumn}
\usepackage{bm}
\usepackage{epsfig}
\usepackage{dsfont,psfrag}
\usepackage{color}
\usepackage{ifthen}

\def\refeq#1{(\ref{#1})}
\begin{document}
\bibliographystyle{amsplain}
  \title{Exponential asymptotic spin correlations in anisotropic
    spin-1/2 XY chains at finite temperatures}
\author{J\"orn Krones}
\author{Joachim Stolze}
\email{joachim.stolze@tu-dortmund.de}
\affiliation{Technische Universit\"at Dortmund, Institut f\"ur Physik, D-44221 Dortmund, Germany}

\date{\today}
\begin{abstract}
The long-time and long-distance asymptotic behavior of the $x$ spin
correlations at finite temperature in an anisotropic
spin-1/2 XY chain is determined numerically. The decay of the
correlations is exponential in both space and time. Similar
exponential decay of correlations was already found earlier in the
special case of the isotropic model, where analytical expressions  for
the decay rates
could be derived via a mapping to a different model. While no such
mapping is known for the anisotropic model, the asymptotic
correlations can be very well approximated by a natural generalization
of the known analytic results for the isotropic case.
\end{abstract}
\maketitle

Spin-1/2 chains are simple quantum many-particle systems which can be
defined in terms of a small number of coupling parameters but
nevertheless offer a rich variety of interesting static and dynamic
phenomena. 
The simple
structure of these models has made possible a large number of links to
other models and fields and has fascinated researchers ever since the
early days of Ising \cite{Isi25} and Bethe \cite{Bet31}.

The XY chain \cite{LSM61,Kat62} is an especially simple example since it can be mapped to
a system of noninteracting lattice fermions. Its dynamics are
nevertheless nontrivial since single-site spin operators are mapped to
strings of Fermi operators and hence two-spin correlation functions
correspond to many-fermion correlation functions which may become
cumbersome to evaluate analytically.

However, numerical calculations have been important in hinting at the
direction where to look for exact analytical results. An early example
is the numerical calculation of Sur et al. \cite{SJL75} for up to nine
spins, suggesting a Gaussian behavior of the infinite temperature $x$
autocorrelation of the isotropic XY chain. That numerical evidence was
soon corroborated by independent rigorous proofs for the general XY
model from two groups \cite{BJ76,BJ77,CP77}. Here we report numerical
results for the long-time and long-distance asymptotic behavior of the
$x$ spin pair correlation functions of anisotropic XY
chains. Analytical results for these correlations were derived for the
special case of the isotropic model only \cite{IIKS93}. Our numerical results
indicate that the analytic formulae valid in the isotropic case
possess natural extensions into the anisotropic regime.

The  $S=1/2$ XY model \cite{LSM61,Kat62} is defined by
the Hamiltonian
\begin{widetext}
\begin{equation}
  \label{ham_spin}
  H= - \sum_{i=1}^{N-1} 
\left\{
J \left[
(1+\gamma) S_i^x S_{i+1}^x + (1-\gamma) S_i^y S_{i+1}^y 
\right]
+h S_i^z
\right\};
\end{equation}
\end{widetext}
with anisotropy parameter $0 \leq \gamma \leq 1$. The limiting cases
$\gamma=0$ and $\gamma=1$ are the isotropic XX and 
transverse Ising (TI) chains, respectively. The Jordan-Wigner
transformation \cite {JW28,LSM61,Kat62}
\begin{equation}
S_i^z = a_i^{\dag} a_i - \frac{1}{2} \, ,
\label{jw1}
\end{equation}
\begin{equation}
S_i^+ = (-1)^{ \sum_{k=1}^{i-1} a_k^{\dag} a_k } a_i^{\dag} \, , \;\;\;
S_i^- = a_i (-1)^{ \sum_{k=1}^{i-1} a_k^{\dag} a_k } \, ,
\label{jw2}
\end{equation}
between the spin-1/2 operators $S_i^z, S_i^{\pm} = S_i^x \pm i S_i^y$
at lattice sites $i$ and the creation and annihilation operators
$a_i^{\dag}, a_i$ of lattice fermions maps the spin Hamiltonian
(\ref{ham_spin}) to a Hamiltonian of noninteracting fermions:
\begin{widetext}
\begin{equation}
  \label{ham_fermion}
  H=  - \sum_{i=1}^{N-1} 
\left\{
\frac J2 \left[ a_i^{\dag} a_{i+1} + a_{i+1}^{\dag} a_{i}
 + \gamma \left(
a_i^{\dag} a_{i+1}^{\dag} +  a_{i+1} a_{i}
\right)
           \right]
+h \left(
a_i^{\dag} a_{i} - \frac 12
\right)
\right\} .
\end{equation}
\end{widetext}
Note that for $\gamma \neq 0$ the number of fermions is not
conserved. From \refeq{jw1},\refeq{jw2} it is evident that the spin
correlation functions 
$\langle S_i^z(t) S_j^z \rangle$
are essentially fermion density correlation functions, while the 
correlation functions 
$\langle S_i^x(t) S_j^x \rangle$
are much more complicated when expressed in the fermion
representation. The fermionic identity
\begin{equation}
(-1)^{  a_k^{\dag} a_k } = (a_k^{\dag} + a_k)(a_k^{\dag} - a_k)
\label{fermi_identity}
\end{equation}
converts the string of signs appearing in \refeq{jw2} into an
expectation value of $2(i+j-1)$ fermion operators, which may be
expressed as a Pfaffian \cite{my28} by Wick's theorem. Pfaffians are
close relatives of determinants and play a role in several statistical
mechanical problems \cite{GH64}. Their numerical evaluation proceeds
along similar lines as that of determinants. The numerical
calculations of this study were all performed for spin chains with
open boundary conditions. Periodic boundary conditions, while
desirable from an aesthetic point of view, lead to additional boundary
terms in the fermionic model \cite{LSM61,Kat62} which make numerical
calculations awkward, if not impossible. Of course, finite-size and
boundary effects are a matter of concern in any numerical
calculation. They are also a topic of research in their own right and
have been studied earlier \cite{GC80,my18,my28}. In the present
study, however, we focus on the asymptotic behavior of bulk spin
correlations in the thermodynamic limit. To make sure that open-chain
numerical results pertain to that situation, only spins sufficiently
far from the boundaries of sufficiently long chains may be
considered. We have checked that the numerical results to be presented
below are not subject to finite-size or boundary effects during the
time intervals shown.

Several results about exact and asymptotic properties of the dynamic
spin correlation functions have been obtained over the years. The
asymptotic time dependence of the longitudinal correlation function
$\langle S_i^z(t) S_j^z \rangle$
is $\sim t^{-1}$ for all $i$ and $j$ in the bulk of the system at all
temperatures \cite{Nie67,KHS70,MS84}. This can be traced back to the
properties of the one-particle density of states of the Jordan-Wigner
fermions and the fact that 
$\langle S_i^z(t) S_j^z \rangle$
is related to fermion density correlations. For $i$ and $j$ close to
the boundary of a long open chain the situation is a little more
complicated \cite{my18,GC80}. 

Due to its more complicated structure in the fermion representation,
the transverse correlation function
$\langle S_i^x(t) S_j^x \rangle$
is more sensitive to temperature variations. At infinite temperature
it vanishes for $i \neq j$ and shows Gaussian decay for $i=j$
\cite{SJL75,BJ76,BJ77,CP77}.  At zero temperature that correlation function exhibits an
asymptotic power-law decay in both space and time for the isotropic (XX) chain.
\cite{MPS83a,MPS83b}.  For finite temperature and in
the isotropic case Its et al \cite{IIKS93} showed that the decay of 
$\langle S_i^x(t) S_j^x \rangle$
is asymptotically exponential in both space and time; numerical
calculations \cite{my28} could be used to assess the range of
validity of the exponential asymptotics. Similar exponential behavior
at finite temperature was also observed in the TI chain 
 \cite{SY97} and in rather general one-dimensional gapless
integrable models  \cite{KS97}. Asymptotic
finite-temperature correlations of the TI chain were also studied in
\cite{RT06,AKT06}; many interesting results on the zero-temperature
correlations of that model were recently obtained by Perk and Au-Yang
\cite{PA09}.

Before discussing our numerical results it is useful to recapitulate
what is known \cite{TM85} about the ground state of the XY Hamiltonian
\refeq{ham_spin} for general anisotropy $\gamma$. Employing a
Fourier transform followed by a Bogoljubov transform, $H$ can be
brought into diagonal free-fermion form with the single-particle
spectrum 
\begin{widetext}
\begin{equation}
  \label{disp_rel}
  \varepsilon_k = - \mathrm{sign} (h+J \cos k) \sqrt{(h+J \cos k)^2 + \gamma^2
    J^2 \sin^2 k}; \quad (-\pi \leq k \leq \pi)
\end{equation}
\end{widetext}
The spectrum (for $|h| \leq |J|$) has two branches with negative and
positive one-particle energies, respectively, with a gap of size
$\Delta \varepsilon = 2 |\gamma| \sqrt{J^2-h^2}$ at the critical wave
vector given by $\cos k_c = -\frac hJ$. The ground state in the
fermionic picture has all negative-energy states occupied and all
others empty. The spectral gap closes at the critical field strength
$|h|=|J|$ (for arbitrary $\gamma$) and for the isotropic case
$\gamma=0$ and arbitrary $h$ (the XX chain), the case considered by
Its et al \cite{IIKS93}.  For $h\neq 0$ the ground state exhibits
long-range order along the $z$ axis; the magnetization $\langle S_i^z
\rangle$ is known analytically \cite{TM85}. For $\gamma \neq 0$ and
$|h|<|J|$ the ground state also exhibits long-range order in the $xy$
plane \cite{BM71} :
\begin{equation}
  \label{x_lro}
  \lim_{r \to \infty} |\langle S_i^x S_{i+r}^x\rangle|^{\frac 12} = 
  \frac {\left[ \gamma^2 \left(1- \left(\frac hJ \right)^2
      \right)\right]^{\frac 18}}
  {\sqrt{2(1+\gamma)}} .
\end{equation}
At finite temperature the equilibrium spin correlation functions 
$\langle S_i^{\alpha}  S_{i+r}^{\alpha}\rangle$
were shown to decay exponentially with $r$ for all $\alpha=x,y,z$
\cite{BM71}. For the XX chain, $\gamma=0$, and subcritical fields,
$|h|<|J|$, the finite-temperature asymptotics of 
$\langle S_i^{x}(t)  S_{i+r}^{x}(0)\rangle$
were found to be exponential in both space and time
\cite{IIKS93}. The derivation of that result proceeded by mapping the
correlation function to the solution of a classical nonlinear
integrable system related to the nonlinear Schr\"odinger equation. 
Despite that intricate derivation the result can be simply interpreted
\cite{IIKS93} in terms of the free energy of quasiparticles with
dispersion (\ref{disp_rel}) (for $\gamma=0$). Generalizing to $\gamma \neq 0$,
we conjecture the following asymptotic formula:
\begin{equation}
\langle S_i^x(t)S_{i+n}^x\rangle \; \; \propto \; \; \left\{
\begin{array}{cc}
\exp \left( - f(n,0) \right) \, , \; \; & n/v_{0}t > 1 \\
Ct^{4\nu^2} \exp \left( - f(n,t) \right) \, , & n/v_{0}t < 1
\end{array}
\right. .
\label{conjecture}
\end{equation}
where
\begin{equation}
f(n,t) = \frac{1}{2 \pi} \int_{-\pi}^{\pi} dk \, \left| n - t \frac{d
    \varepsilon_k}{dk}  \right| \,
 \ln  \left|\tanh
\frac{\beta \varepsilon_k}{2}\right|  \, ,
\label{exponent}
\end{equation}
and where:
\begin{equation}
\nu=\frac{1}{2\pi}\ln\left| \tanh\left(\frac{\beta\varepsilon_{k_0}}{2}\right)\right| ,
\end{equation}
while $k_0$ is determined by $\frac{d}{dk}\varepsilon_{k_0}=\frac{n}{t}$. Furthermore,
\begin{equation}
  \label{v0}
  v_{0} = \max_{|k| \leq \pi} \left| \frac{d \varepsilon_k}{dk}   \right|
\end{equation}
is the maximum value of the group velocity at which the fermionic quasiparticles move.
In the spacelike region, $n>v_{0}t$, the correlation is 
constant and equal to its stationary value. This is due to the fact
that the two spins at distance $n$ have not been able to communicate
with each other yet. Only later can the fermionic
quasiparticles of energy $\varepsilon_k$ transfer any information at
group velocity $\frac{d \varepsilon_k}{dk}$ between the two spins. 
For zero magnetic field, $h=0$, the maximum velocity $v_{0}$ turns out \cite{Noe}
 to be 
\begin{equation}
  \label{v0h0}
  v_{0} = (1-\gamma) J  ;
\end{equation}
for nonzero field there is no simple closed expression for $v_{0}$, but numerical evaluation is simple.

\begin{figure}[h]
\includegraphics[width=\columnwidth,clip]{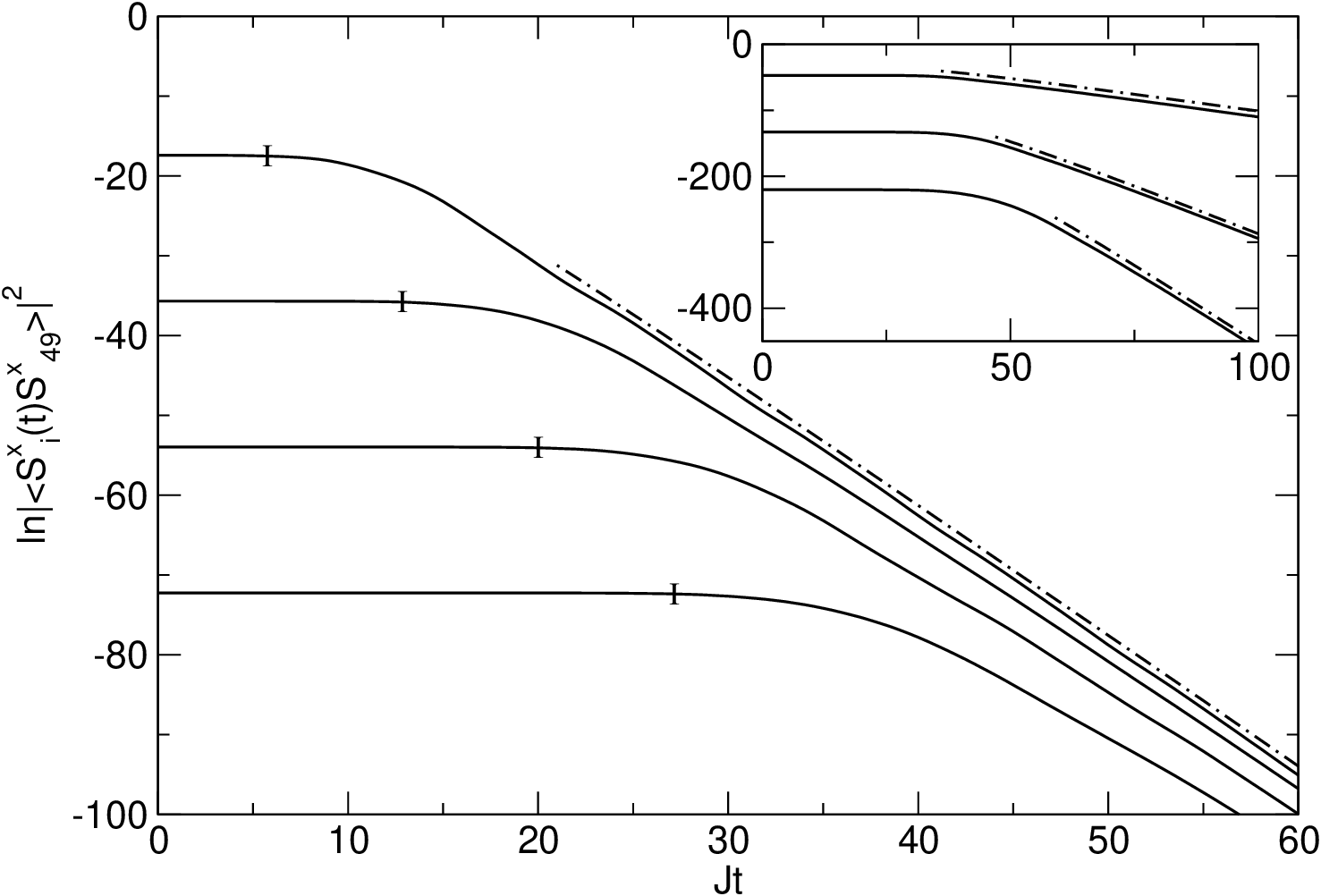}
\caption{Main plot: Spin pair correlation function $\langle S_i^x(t)S_j^x\rangle$
in the bulk regime of the anisotropic XY chain, $\gamma=0.3$, without external field, $h=0$,
at $T/J=2$, for $j=49$ and $i=30, 35, 40,$ and $45$
(solid lines, bottom to top) in an open chain with $N=100$  spins.
Plotted is the logarithm of $|\langle S_i^x(t)S_j^x\rangle|^2$.
The marks on the curves denote the boundary between space-like and time-like regions, compare \refeq{conjecture}.
The inset shows the same function in the same representation, but here the
sites are kept fixed $(i=30 , j=49)$ and the temperature is varied
($T/J=1, 10, 100$, top to bottom).
The slopes of the dot-dashed straight-line segments represent the decay rate
inferred from the asymptotic expression (\ref{conjecture}).}
\label{eins}
\end{figure}

The
crossover between spacelike and timelike regions is illustrated in
Fig.~\ref{eins}. 
In that figure we show a logarithmic
plot of $|\langle S_i^x (t)S_{i+n}^x\rangle|^2$
for $n$=4, 9, 14, 19 at temperature $T/J=2$ and anisotropy $\gamma=0.3$.
We observe that each function is almost perfectly constant up to the time
$v_0 t_n= n$, where it bends smoothly into exponential decay with weak
superimposed oscillations.
The decay time does not show any significant dependence on $n$.
The inverse decay time predicted by (\ref{conjecture}) for the asymptotic
regime of the uppermost curve is given by the slope of the dot-dashed line
and matches our data very well.
The linear variation with $n$ of the intercepts at $t=0$ in this
logarithmic plot reflects the well-established exponential decay of the
equal-time correlation function
$\langle S_i^x S_{i+n}^x \rangle \sim \exp[-n/\xi(T)]$.
\par
The inset to Fig. \ref{eins} shows again the curve $n=19$ of the main plot
along with curves for the same correlation function at different temperatures.
Now the crossover between the space-like and the time-like regime occurs
roughly at one common value of $Jt$.
In the time-like regime, the slope changes from one curve to the next, which
reflects the $T$-dependence of the decay time, while the variable intercept
in the space-like regime reflects the $T$-dependence of the correlation
length.

\begin{figure}[h]
\includegraphics[width=\columnwidth,clip]{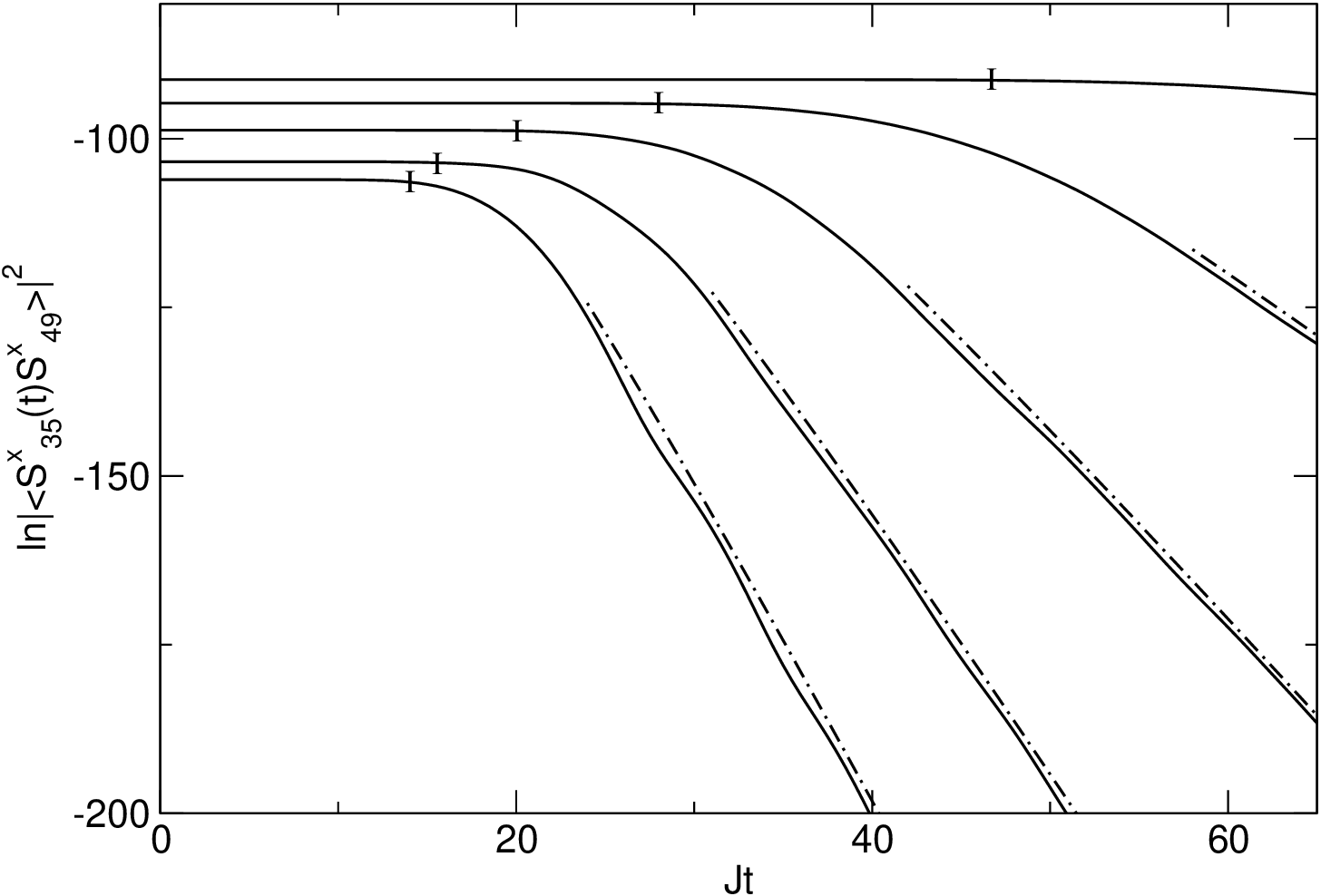}
\caption{
Same as Fig. \ref{eins}, for fixed distance $n=14$ between the two
spins, and fixed temperature, $T/J=10$, at $h=0$. The anisotropy parameter is varied: $\gamma=0, 0.1, 0.3, 0.5, 0.7$ (bottom to top). The marks on the curves denote the boundary between space-like and time-like regions, compare \refeq{conjecture}.
The slopes of the dot-dashed straight-line segments represent the decay rate
inferred from the asymptotic expression (\ref{conjecture}).
}
\label{zwei}
\end{figure}

In Fig. \ref{zwei} we show results for fixed spatial distance $n=14$
and fixed temperature  $T/J=10$, for varying anisotropy $\gamma$. For
growing $\gamma$ the decay time in the asymptotic regime grows, as
does the crossover time between the space-like and time-like
regimes. For $\gamma \to 1$ the crossover time diverges since the
velocity $v_0$ (\ref{v0h0}) vanishes. The decay time also diverges as
$\gamma \to 1$. This is to be expected, since in the absence of a
magnetic field the model at $\gamma = 1$ (i.e. the transverse Ising
chain) does not display any dynamics. The $\gamma$-dependence of the
correlation length can be read off from the $t=0$ intercepts of the
curves shown in Fig. \ref{zwei}.

\begin{figure}[h]
\includegraphics[width=\columnwidth,clip]{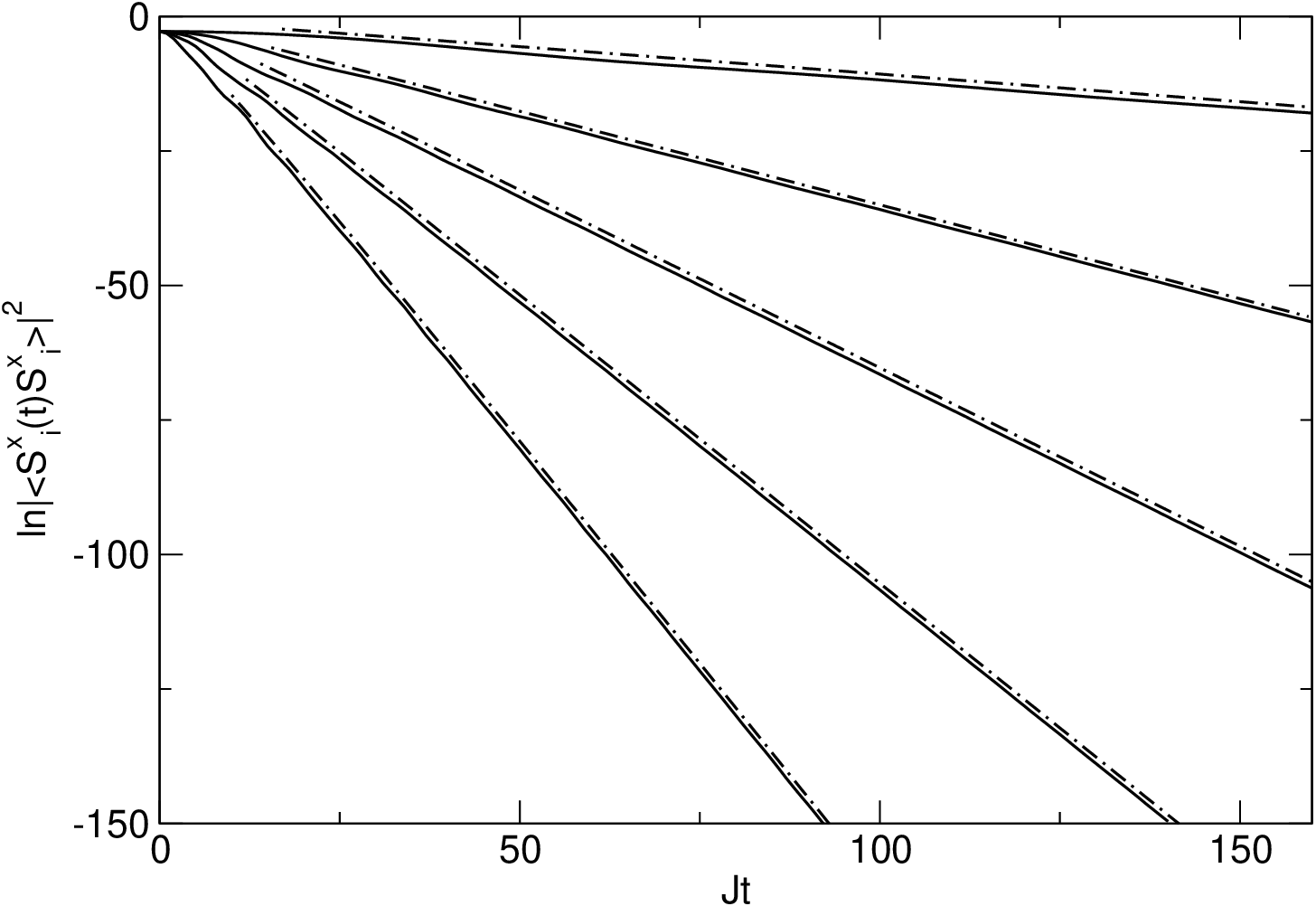}
\caption{
Spin autocorrelation function $\langle S_i^x(t)S_i^x\rangle$
in the bulk regime of the anisotropic XY chain, at $h=0$,
$T/J=1$, for $i=100$  in an open chain with $N=200$ spins.
Plotted is the logarithm of $|\langle S_i^x(t)S_i^x\rangle|^2$.
The anisotropy values are (bottom to top) $\gamma=0.1, 0.3, 0.5, 0.7$,
and 0.9.
The slopes of the dot-dashed straight-line segments represent the decay rate
inferred from the asymptotic expression (\ref{conjecture}).
}
\label{drei}
\end{figure}

Fig.  \ref{drei} shows the autocorrelation function $(n=0)$ over a
longer time interval, demonstrating very clearly how well the
conjectured formula (\ref{conjecture}) fits the numerical results.

\begin{figure}[h]
\includegraphics[width=\columnwidth,clip]{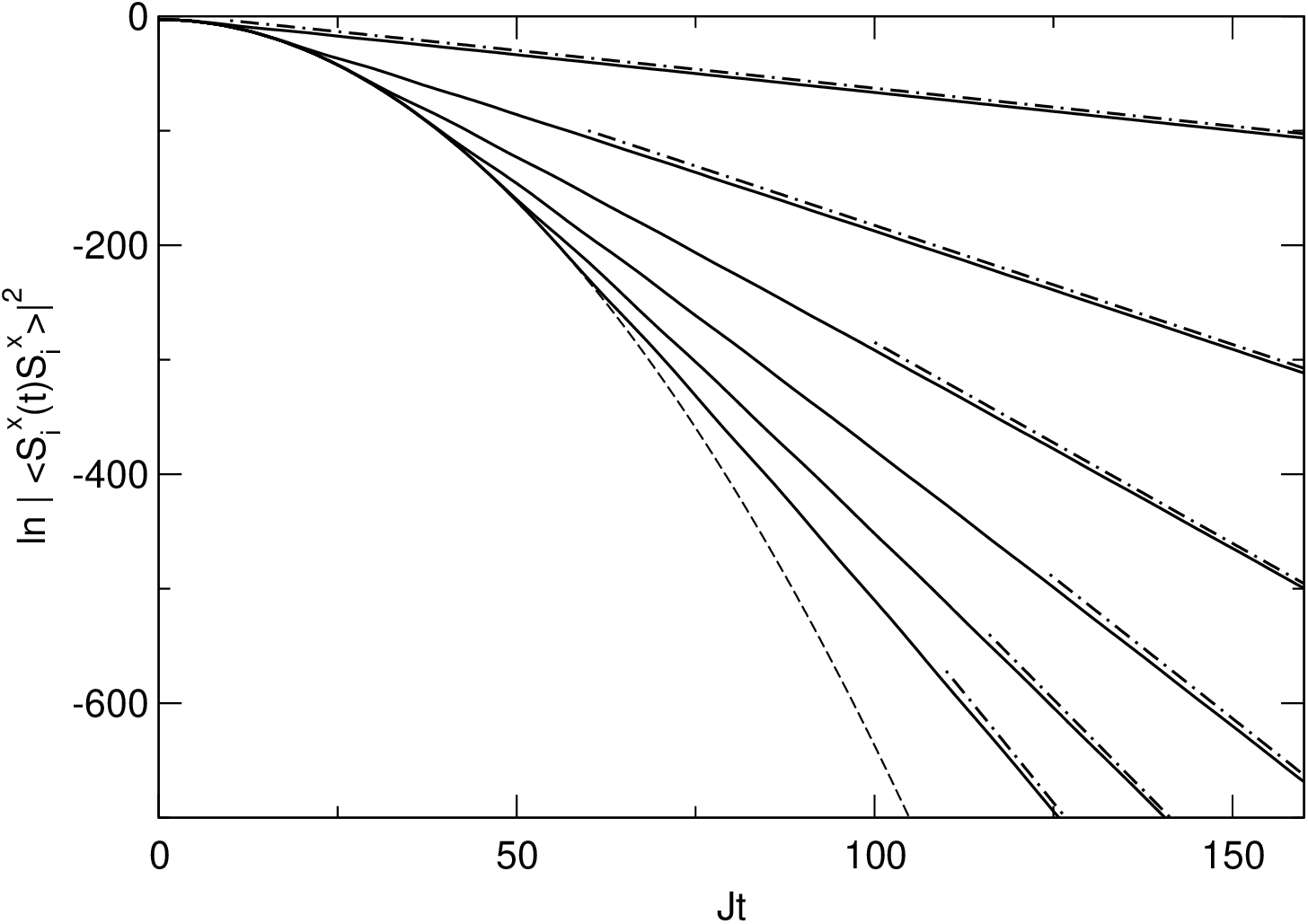}
\caption{
Same as Fig. \ref{drei}, for fixed anisotropy $\gamma=0.5$ and varying
temperature $T=1, 10, 100, 1000, 10^4$, and $10^5$ (top to bottom).
}
\label{vier}
\end{figure}

The temperature dependence of the autocorrelation function is
displayed in Fig. \ref{vier} for fixed anisotropy, $\gamma=0.5$. Two
regimes can be clearly distinguished: at short times the
autocorrelation is a Gaussian, crossing over to an exponential
behavior at longer times. The crossover time seems to grow roughly
logarithmically with temperature, so that at $T \to \infty$ the
well-known Gaussian behavior \cite{BJ76,BJ77,CP77} emerges.

The dependence of the spin correlation function on the external field
was also observed to follow the asymptotic formula (\ref{conjecture})
for subcritical ($|h|<|J|$) field values.

To conclude, we have found convincing numerical evidence for
exponential decay of the finite-temperature $x$ spin correlations of
anisotropic XY chains in both space and time. We suggest an analytic
formula (\ref{conjecture}) for that exponential decay which fits the
numerical data well and which generalizes the known analytical result
\cite{IIKS93} based on a mapping to a classical nonlinear integrable
system. At high temperature our numerical results connect well to the
known \cite{BJ76,BJ77,CP77} Gaussian infinite-temperature behavior.

\newcommand{\noopsort}[1]{}

\end{document}